\pdfoutput=1

\documentclass[11pt]{article}

\usepackage{EMNLP2022}

\usepackage{times}
\usepackage{latexsym}
\usepackage{adjustbox}
\usepackage{amsmath}

\usepackage[T1]{fontenc}

\usepackage[utf8]{inputenc}

\usepackage{microtype}

\usepackage{inconsolata}

\usepackage{microtype}
\usepackage{algorithm}
\usepackage{algorithmic}
\usepackage{graphicx}
\usepackage{color}
\usepackage{booktabs}
\usepackage{multirow}
\usepackage{verbatim}
\newcommand{\paratitle}[1]{\vspace{1.5ex}\noindent\textbf{#1}~}
\newcommand{\datatitle}[1]{\noindent\textbf{#1}}
\newcommand{\gmodel}{GNN-encoder~}{}
\newcommand{\modelnospace}{GNN-encoder}{}
\newcommand{\ie}{\emph{i.e.,}}

\title{GNN-encoder: Learning a Dual-encoder Architecture via Graph Neural Networks for Dense Passage Retrieval}

\author{Jiduan Liu$^{1,2}$\thanks{\llap{}\:\:\:Work done during internship at Meituan.}\ \ , Jiahao Liu$^{3}$, Yang Yang$^{3}$, Jingang Wang$^{3}$, Wei Wu$^{3}$ \\
\textbf{Dongyan Zhao$^{1,4,5}$$^\dagger$, Rui Yan$^{6}$}\thanks{$^\dagger$ Corresponding authors: Dongyan Zhao
(zhaody@pku.edu.cn) and Rui Yan (ruiyan@ruc.edu.cn).} \\
$^1$Wangxuan Institute of Computer Technology, Peking University \\
$^2$Center for Data Science, AAIS, Peking University; $^3$Meituan \\
$^4$Key Laboratory of Computational Linguistics (Peking University), Ministry of Education\\
$^5$Beijing Institute for General Artificial Intelligence \\
$^6$Gaoling School of Artificial Intelligence, Renmin University of China \\
\texttt{\{liujiduan, zhaody\}@pku.edu.cn}, \texttt{wuwei19850318@gmail.com}, \texttt{ruiyan@ruc.edu.cn}\\
\texttt{\{liujiahao12, yangyang113, wangjingang02\}@meituan.com}}

\begin{document}
\maketitle
\begin{abstract}
Recently, retrieval models based on dense representations are dominant in passage retrieval tasks, due to their outstanding ability in terms of capturing semantics of input text compared to the traditional sparse vector space models. A common practice of dense retrieval models is to exploit a dual-encoder architecture to represent a query and a passage independently. Though efficient, such a structure loses interaction between the query-passage pair, resulting in inferior accuracy. To enhance the performance of dense retrieval models without loss of efficiency, we propose a GNN-encoder model in which query (passage) information is fused into passage (query) representations via graph neural networks that are constructed by queries and their top retrieved passages. By this means, we maintain a dual-encoder structure, and retain some interaction information between query-passage pairs in their representations, which enables us to achieve both efficiency and efficacy in passage retrieval. Evaluation results indicate that our method significantly outperforms the existing models on MSMARCO, Natural Questions and TriviaQA datasets, and achieves the new state-of-the-art on these datasets.

\end{abstract}
\pdfoutput=1
\section{Introduction}
Large-scale query-passage retrieval is a core task in search systems, which aims to rank a collection of passages based on their relevance with regard to a query. To balance efficiency and effectiveness, existing work typically adopts a two-stage retrieval pipeline~\cite{renrocketqav2,twostageQA}. The first-stage aims to retrieve a subset of candidate passages by a recall model from the entire corpus and the second stage aims to re-rank the retrieved passages. 
In the first-stage retrieval, traditional approaches~\cite{traditionalqa} implemented term-based retriever (e.g. TF-IDF and BM25) by weighting terms based on their frequency, which have limitations on representing semantics of text. Recently, dense passage retrieval is drawing more and more attention in the task of passage retrieval~\cite{karpukhin2020dense}. The underlying idea is to represent both queries and passages as embeddings, so that the semantic relevance can be measured via embeddings similarity. 
With the great success of pre-trained language models (PLMs) such as BERT/RoBERTa~\cite{BERT,RoBERTa} in natural language processing tasks, dense retrieval models parameterized by PLMs is emerging as the new state-of-the-art in a variety of passage retrieval tasks \cite{karpukhin2020dense,xiong2020approximate}. 

Two paradigms based on fine-tuned language models are typically built for retrieval: cross-encoders and dual-encoders. 
Typical cross-encoders need to recompute the representation of each passage in the corpus once a new query comes, which is difficult to deploy in real-world search systems. 
In contrast, dual-encoders remove query-passage interaction by representing a query and a passage independently through two separate encoders (Siamese encoders). Hence, passage embeddings can be pre-computed offline, and online latency can be greatly reduced.
Thanks to this advantage, dual-encoders are more widely adopted in real-world applications. 
On the other hand, independent encoding without any interaction causes severe retrieval performance drop due to information loss. To improve the performance of dual-encoders, some efforts have been made to incorporate more complicated structures (i.e., late interaction) such as attention layers~\cite{Humeau2019polyencoder, tang2021Pseudo}, the sum of maximum similarity computations ~\cite{khattab2020colbert}, and the transformer layers~\cite{deformer, dipair} into encoding. 
These late interaction strategies bring considerable improvements on retrieval performance but also increase computational overhead. Moreover, interaction information is still neglected in earlier encoding of query and passage.

In this work, we aim to achieve both efficiency and effectiveness in passage retrieval. The key idea is to maintain two independent encoders, and keep as much interaction information as possible in the meanwhile. To this end, we propose a novel approach that explicitly fuses query (passage) information into passage (query) embeddings through a graph neural network (GNN), and name the model \modelnospace. Our model is built upon the dual-encoder, and learns query-interactive passage representations and passage-interactive query representations through a graph neural network. Specifically, given a query set, we retrieve top passages for each query, and form a graph whose nodes are the queries and the passages, and edges reflect correspondence between query-passage pairs (i.e., if a passage is retrieved by the query). 
Then, we initialize the GNN model with the representations of the pre-trained dual-encoder and cross-encoder, and then perform information propagation on the graph.
To avoid information leakage, we further design a new training algorithm and name it Masked Graph Training (MGT), in which the query set used for training GNN is no longer used to construct the query-passage graph in each training epoch. 
Finally, the passage embeddings could be pre-computed offline corresponding to the GNN. 
Thus our model holds the efficiency advantage inherited from the dual structure, and at the same time takes query-passage interaction into account.


Our contributions can be summarized as follows:
\begin{itemize}
    \item We propose a novel dense retrieval model based on graph neural network techniques that encodes query-passage interaction into query and passage representations without sacrifice on retrieval efficiency.  
    \item We propose an adaptive Masked Graph Training (MGT) Algorithm for our task to avoid information leakage during GNN training.
    \item Experiments show that our model achieves state-of-the-art performance on MSMARCO, Natural Questions and TriviaQA datasets.
\end{itemize}
\pdfoutput=1
\section{Related Work}
Our work touches on two strands of research within dense passage retrieval and graph neural network. 

\subsection{Dense Passage Retrieval}

The dense passage retrieval approaches have been proposed to map both questions and documents to continuous vectors (i.e., embeddings), which has achieved better performance than sparse retrieval approaches~\cite{traditionalqa, dai2019deepct}.
Existing approaches can be roughly divided into two categories: pre-training and fine-tuning. The first type of methods often explores pre-training objectives/architectures designed for retrieval.
~\citet{lee2019ICT} pre-trains the retriever with an unsupervised Inverse Cloze Task (ICT). Condenser~\cite{gao2021condenser} proposes a dense retrieval pre-training architecture which learns to condense information into the dense vector through LM pre-training. coCondenser~\cite{gao2021unsupervised} adds an unsupervised corpus-level contrastive
loss on top of the Condenser~\cite{gao2021condenser} to warm up passage embeddings.

The second type of methods often fine-tunes pre-trained language models on labeled data. ~\citet{karpukhin2020dense} proposes a dense embedding model using only pairs of questions and passages, without additional pre-training. \citet{xiong2020approximate, qu2021rocketqa} identify that the negative samples during training may not be representative, thus mechanism of selecting hard training negatives is designed. ~\citet{khattab2020colbert, Humeau2019polyencoder} incorporate late interaction architectures into the learning process that independently encode the query and the document firstly. 
\citet{tang2021Pseudo} 
designs a method to mimic the queries on each of the documents by clustering to enhance the document representation.
PAIR \cite{ren2021pair} leverages passage-centric similarity relation into training object to discriminate between positive and negative passages.
RocketQAv2 \cite{renrocketqav2} introduces dynamic listwise distillation to jointly train retriever and re-ranker.

\begin{figure*}
    \centering
    \includegraphics[width=2.0\columnwidth]{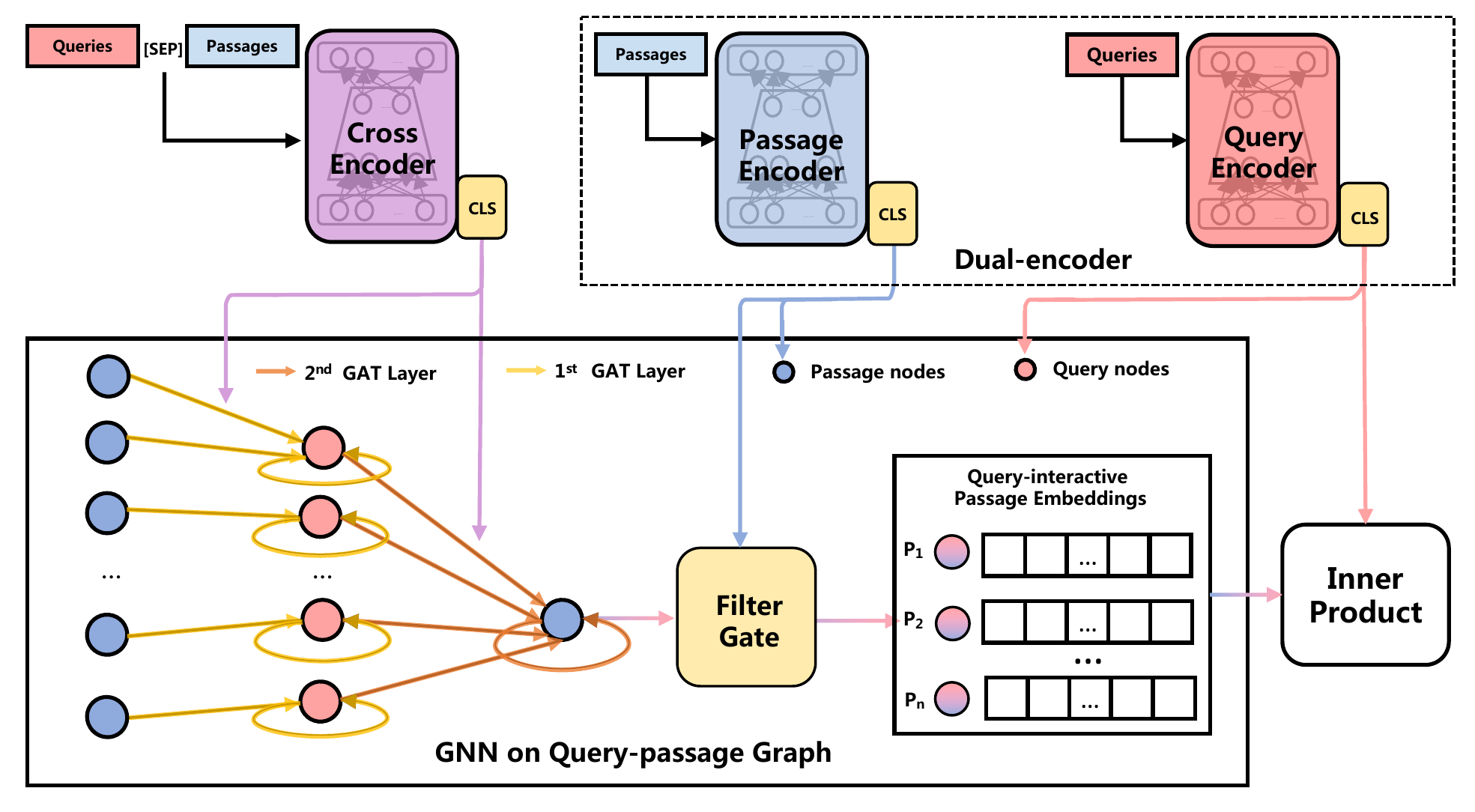}
    \caption{
        Overview of \gmodel which can be divided into three parts: (1) Dual-encoder; (2) Cross-encoder; (3) GNN on query-passage graph. Node and edge features of query-passage graph are initialized by dual-encoder and cross-encoder, respectively. Only the parameters of dual-encoder and GNN will be updated during training.
    }
    \label{fig:model}
\end{figure*}

\subsection{Graph Neural Network}
Graph neural network (GNN) captures the relationships between nodes connected with edges, which propagates features across nodes layer by layer~\cite{scarselli2008gnn}. Graph attention network (GAT)~\cite{velivckovic2018graph} leverages masked self-attentional layers to address the shortcomings of prior work based on graph convolutions or their approximations. GNN has demonstrated effectiveness in a wide variety of tasks such as text classification \cite{lin2021bertgcn}, question answering ~\cite{de2018qagcn},  recommendation~\cite{wu2019recommendationgcn} and relation extraction~\cite{li2020REgraph}. For example, 
~\citet{QAGNN} connects the question-answering context and knowledge graph to form a joint graph, and mutually updates their representations to perform joint reasoning over the language and the knowledge graph. 

Compared to existing work, our work serves the dense passage retrieval and presents a novel fine-tuning method to fuse query (passage) information into passage (query) embeddings via GNN. 

\pdfoutput=1
\section{Methodology}
\subsection{Preliminary}
Given a query $q$, dense retriever is required to retrieve k most relevant passages $\{p_i\}_{i=1}^k$ from a large corpus consisting of hundreds of thousands of passages. For the sake of retrieval efficiency, the dual-encoder architecture is widely adopted, where query encoder $E_Q(\cdot)$ and passage encoder $E_P(\cdot)$ are used to embed query $q$ and passage $p$ into $d$-dimensional vectors, respectively. The similarity between query $q$ and passage $p$ can be computed as the dot product of their vectors:
\begin{equation}\label{eq:sim}
    s(q,p) = E_Q(q)^\mathrm{T}\cdot E_P(p).
\end{equation}

The training objective of the dual-encoder is to learn embeddings of queries and passages to make positive query-passage pairs have higher similarity than the negative query-passage pairs in training data. Hence, the contrastive-learning loss function is adopted for the dual-encoder:
\begin{equation}\label{loss}
     \begin{aligned}
    &L(q,p^+,\{p^-\},s) = \\
    &- \log \frac{e^{s(q,p^+)}}{e^{s(q,p^+)}+\sum_{\{p^-\}}e^{s(q,p^-)}},
    \end{aligned}
\end{equation}
where $q$ and $p^+$ represent query and positive passage, respectively, and $\{p^-\}$ represents the set of negative passages.

In practical retrieval systems, passage embeddings are usually pre-computed offline, while query embeddings are computed by the query encoder in an ad hoc manner. Therefore we can obtain better passage embeddings through a complicated encoder as long as it does not increase the online inference latency.

\subsection{Graph Construction} \label{3.2} We use all the passages $P = \{p_i\}_{i=1}^m$ and training queries 
$Q = \{q_i\}_{i=1}^n$ as nodes to construct our \textbf{query-passage graph}, where $n$ and $m$ denote the number of training queries and passages, respectively. Let $\mathcal{G} = (\mathcal{V},\mathcal{E})$ denote our graph, where $\mathcal{V} = \mathcal{V}_q \cup \mathcal{V}_p$ is a node set and $\mathcal{E} = \mathcal{E}_{pp} \cup \mathcal{E}_{pq} \cup \mathcal{E}_{qq}$ is an edge set. 
$\mathcal{V}_p$ and $\mathcal{V}_q$ denote the passage node set and the query node set, respectively. $\mathcal{E}_{pp}$, $\mathcal{E}_{pq}$, and $\mathcal{E}_{qq}$ denote the edges between passage nodes, between passage nodes and query nodes, and between query nodes, respectively. 
The edge between node $x$ and node $y$ is denoted as $e(x,y)$.

We retrieve the top-k candidate passages $P_i = \{p_{ij}\}_{j=1}^k$ for each query $q_i$ from the corpus by the dual-encoder. We add edges between $q_i$ and each passage $p_{ij}$ in the top-k retrieved passages $P_i$ which may be relevant to $q_i$. These edges compose $\mathcal{E}_{pq}$ (i.e., $\mathcal{E}_{pq} = \{e(q_i,p_{ij})\}_{i=1,j=1}^{n,k}$). Since we can not directly distinguish whether there is relation between the queries and between the passages, we only add self-loops to each passage and query (i.e., $\mathcal{E}_{pp}=\{e(p_i,p_i)\}_{i=1}^m$ and $\mathcal{E}_{qq}=\{e(q_i,q_i)\}_{i=1}^n$). To summarize, our graph $\mathcal{G}$ has a total of $(n+m)$ nodes and $(n\times k + m +n)$ edges.

\paratitle{Node Features} 
We use the dual-encoder to get query embeddings $h_{q_i} = E_Q(q_i)$ and passage embeddings $h_{p_i} = E_P(p_i)$ as our graph node features. 

\paratitle{Edge Features} 
Since the cross-encoder can better capture the interactive information between text pairs, we utilize the embeddings of text pairs $(x,y)$ by the cross-encoder as features $h_{x-y}$ of edge $e(x,y)$. We believe they will guide GNN model to choose information of neighbor nodes. 

In our experiments, both dual-encoder and cross-encoder use [CLS] representations as embeddings.

\subsection{GNN on Query-passage Graph} \label{3.3}
The most straightforward way to fuse query (passage) information into passage (query) embeddings is directly adding query (passage) embeddings to passage (query) embeddings. However, not all information is effective. Ideally, a model can choose what information to utilize and how much such information should be retained, which is exactly what our proposed model (Figure~\ref{fig:model}) does. Since GAT \cite{velivckovic2018graph} can learn the attention weights to neighbors (i.e., how to choose information of neighbors), it suits well to our work. In our graph, two nodes connected are considered to be relevant, so we can utilize GAT to learn how to exchange information between relevant nodes. 

\paratitle{GAT Layer}
We apply multi-head attention in GAT layer. But for simplicity, we only describe the single-head situation below. Let $\mathcal{N}_i$ denote the neighbors of node $i$ in the graph. GAT layer computes the importance of node $j \in \mathcal{N}_i$ to node $i$ as:
\begin{equation}
    e_{ij} = a^\mathrm{T}[W_t h_i||W_s h_j || W_e h_{i-j}],
\end{equation}
where $h_i \in \mathcal{R}^{d_n}$, $h_j \in \mathcal{R}^{d_n}$, $h_{i-j} \in \mathcal{R}^{d_e}$ are the features of node $i$, node $j$ and edge $e(i,j)$, respectively, and $W_t \in \mathcal{R}^{d_n \times d}$, $W_s \in \mathcal{R}^{d_n \times d}$, $W_e \in \mathcal{R}^{d_e \times d}$, $a \in \mathcal{R}^{3d}$ are learnable model parameters, and $||$ is concatenation operation. In our experiments, we set $d_n = d_e = d$ which equals to the dimension of BERT base (\ie 768). Then the attention weight of node $j \in \mathcal{N}_i$ to node $i$ is calculated by the softmax function and LeakyReLU activation function:
\begin{equation}\label{eq:atention}
    \alpha_{ij} = \frac{\exp(\mbox{LeakyReLU}(e_{ij}))}{\sum_{k \in \mathcal{N}_i}\exp(\mbox{LeakyReLU}(e_{ik}))}.
\end{equation}
The final output features for every node can be computed as weighted sum of linear transformed features of neighbor nodes, and optionally adding an activation function $\sigma$:
\begin{equation}\label{eq:gatOutput}
    \widetilde{h_i} = \sigma(\sum_{j\in \mathcal{N}_i}\alpha_{ij}W_s h_j).
\end{equation}

Considering only node features will change after GAT layer, we define the above formulation as: $\widetilde{h_i} = \mbox{GAT}(\{h_j\}_{j \in \mathcal{N}_i})$ for notation simplicity. We use two GAT layers to implement the interaction between nodes, because two-hop neighbors can exactly establish the relation between the queries and between the passages.

\paratitle{Aggregate}
We first get the aggregated contexts of query neighbors via the first GAT layer:
\begin{equation}\label{eq:gat1}
    \widetilde{h_{q_i}} = \mbox{GAT}(\{h_{p_{ij}}\}_{p_{ij} \in P_i} \cup \{h_{q_i}\}),
\end{equation}
where $P_i$ denotes the set of passages retrieved by $q_i$. Then we concatenate the aggregated contexts and query node features, and apply a single linear layer to get \textbf{passage-interactive query embeddings}:
\begin{equation}\label{eq:iqe}
    h_{q_i}' = W_{pq}[\widetilde{h_{q_i}}||h_{q_i}] + b_{pq},
\end{equation}
where $W_{pq} \in \mathcal{R}^{d\times 2d}$ and $b_{pq} \in \mathcal{R}^d$ are weight matrix and weight vector, respectively. Then we get the aggregated contexts of passage neighbors via the second GAT layer:
\begin{equation}
    \widetilde{h_{p_i}} = \mbox{GAT}(\{h_{q_{ij}}'\}_{{q_{ij}} \in Q_i} \cup \{h_{p_i}\}),
\end{equation}
where $Q_i$ denotes the set of queries retrieving $p_i$. 

\paratitle{Filter Gate}
We can not directly incorporate aggregated contexts of passage neighbors into passage embeddings for the noise in them. We utilize filter gate to "clean"\footnote{Since the passage-interactive query embeddings are not directly involved in the similarity calculation, we should keep as much information as possible for the subsequent calculation. Therefore we do not apply the gate structure to filter noise in the passage-interactive query embeddings, and the final gate will filter all the noise.} the aggregated contexts: 
\begin{equation}\label{eq:gate}
    f_{p_i} = \sigma(W_{qp}[\widetilde{h_{p_i}}||h_{p_i}] + b_{qp}),
\end{equation}
where $W_{qp} \in \mathcal{R}^{d\times 2d}$ and $b_{qp} \in \mathcal{R}^d$ are weight matrix and weight vector, respectively and $\sigma$ represents the sigmoid function. We get the final \textbf{query-interactive passage embeddings} as follows:
\begin{equation}
\label{eq:ipe}
    h_{p_i}' = f_{p_i} \cdot \widetilde{h_{p_i}} + h_{p_i}.
\end{equation}
\subsection{Training Procedure} \label{3.4}
In this section, we present the training procedure of our model. Following \citet{qu2021rocketqa} and \citet{ren2021pair}, we first retrieve the top-k candidates of each query from the corpus by DPR~\cite{karpukhin2020dense} and score them by a well-trained cross-encoder $M_{ce}$ to obtain denoised positives and hard negatives to train our initial dual-encoder $M_{de}$. We utilize $M_{de}$ and $M_{ce}$ to construct our query-passage graph, including adding edges, initializing node features and initializing edge features.

We adopt Eq.(\ref{loss}) as our basic loss function to train the dual-encoder (initialized with $M_{de}$) and GNN jointly. For each training step, we randomly sample a subset of training queries $Q_b = \{q_i\}_{i=1}^b$, getting their denoised positives and hard negatives $P_b = P_b^+ \cup P_b^- = \{p_{q_i}^+\}_{i=1}^b \cup \{p_{q_i}^-\}_{i=1}^b$. We compute query-interactive passage embeddings for each passage in $P_b$, and then utilize them and query embeddings to compute similarity and loss:
\begin{gather}\label{eq:gnnsim}
    s_{\mathcal{G}}(q,p) = E_Q(q)^\mathrm{T} \cdot h_p', \\
    L_{\mathcal{G}} = \sum_{q_i \in Q_b}L(q_i,p^+_{q_i},P_b - \{p^+_{q_i}\},s_{\mathcal{G}}). \label{eq:gnnloss}
\end{gather}
In this way, query encoder can learn how to produce query embeddings with interaction information by being involved in the computation of passage-interactive query embeddings and similarity.

Ideally, we need to recompute the node features by dual-encoder at each training step. However, considering efficiency and hardware resources, we utilize cache passage embeddings of $M_{de}$ as passage node features to compute $\widetilde{h_{q_i}}$ in Eq.(\ref{eq:gat1}), while node features in Eq.(\ref{eq:iqe}$\sim$\ref{eq:ipe}) are recomputed by dual-encoder. For retrieval, we utilize the similarity calculated by Eq.(\ref{eq:gnnsim}) as score to rank all passages. 

\paratitle{Masked Graph Training Algorithm} The method presented above is highly susceptible to information leakage in training. Dropping labeled edges is a common trick to avoid leakage information, but it does not suit our task which will be analysed in later experiments. The training queries are involved in graph construction, while test queries are not, which also leads to the gap between training and inference. When constructing graph, positive passage $p_{q_i}^+$ tends to be retrieved by query $q_i$, which may lead $p_{q_i}^+$ focus too much on $q_i$. Based on these considerations, we propose a Masked Graph Training (MGT) Algorithm which is applicable to other tasks suffering the same dilemma. In every epoch, we split the training queries into two parts $Q_g$ and $Q_t$ which are utilized for constructing the graph and training (i.e., masking some nodes in training), respectively. By this means, we alleviate the gap between training and inference . The proportion of query masked $\beta$ in the graph can not be too high, otherwise it will cause the gap between training graph and inference graph. For more details, you can refer to appendix \ref{appendix:algorithm}.
\pdfoutput=1
\section{Experiments}
\begin{table*}[htp]
\centering
  \begin{tabular}{p{3cm}lccccccc}
    \toprule
    \multirow{2}{*}{\textbf{Methods}} & \multirow{2}{*}{\textbf{PLM}} & \multicolumn{3}{c}{\textbf{MSMARCO Dev}} &
    \multicolumn{3}{c}{\textbf{Natural Questions Test}} &\\
    & & MRR@10 & R@50 & R@1000 & R@5 & R@20 & R@100 \\
    \midrule
    BM25~(anserini) & - & 18.7 & 59.2 & 85.7 & - & 59.1 & 73.7\\
    \midrule
    DeepCT & - & 24.3 & 69.0 & 91.0 & - & - & - \\
    GAR & - & - & - & - & - & 74.4 & 85.3 \\
    COIL & - & 35.5 & - & 96.3 & - & - & - \\
    \midrule
    DPR (single) & BERT$_{\rm base}$ & - & - & - & - & 78.4 & 85.4 \\
    ANCE (single) & RoBERTa$_{\rm base}$ & 33.0 & - & 95.9 & - & 81.9 & 87.5 \\
    ColBERT  & BERT$_{\rm base}$ & 36.0 & 82.9 & 96.8 & - & - & - \\
    NPRINC & BERT$_{\rm base}$ & 31.1 & - & 97.7 & 73.3 & 82.8 & 88.4 \\
    RocketQA  & ERNIE$_{\rm base}$ & 37.0 & 85.5 & 97.9 & 74.0 & 82.7 & 88.5  \\
    PAIR  & ERNIE$_{\rm base}$ & 37.9 & 86.4 & 98.2 & 74.9 & 83.5 & 89.1 \\
    Condenser$^\dagger$ & - & 36.6 & - & 97.4 & - & 83.2 & 88.4 \\
    RocketQAv2  & ERNIE$_{\rm base}$ & 38.8 & 86.2 & 98.1 & 75.1 & 83.7 & 89.0\\
    coCondenser$^\dagger$ & - & 38.2 & - & \textbf{98.4} & 75.8 & 84.3 & 89.0 \\
    \midrule
   \gmodel & ERNIE$_{\rm base}$ & \textbf{39.3} & \textbf{86.9} & 98.3 & \textbf{76.8} & \textbf{84.9} & \textbf{89.3} \\
  \bottomrule
\end{tabular}
 \caption{Experimental results on MSMARCO dev set and Natural Questions test set. We copy the results from original papers, while leaving it blank if unavailable. The best results are marked bold. $^\dagger$Note that Condenser and coCondenser are pre-training methods.}
\label{tab:mainR}
\end{table*}

\subsection{Dataset}
\datatitle{MSMARCO}~\cite{nguyen2016ms} is the dataset introduced by Microsoft. 
It contains 0.5 million queries that were sampled from Bing search logs, while containing 8.8 million passages that were gathered from Bing's results to real-world queries.

\datatitle{Natural Questions} (NQ)~\cite{kwiatkowski2019natural} is a large dataset from open-domain QA, which consists of queries that were issued to the Google search engine by real anonymized, and the collection of passages is processed from Wikipedia.

\datatitle{TriviaQA} (TQA)~\cite{joshi2017triviaqa} contains a set of trivia questions with answers which were originally scraped from the Web.

For both Natural Questions and TriviaQA, we reuse the version released by DPR~\cite{karpukhin2020dense} in our experiments. 
Following previous work, we use mean reciprocal rank (MRR) and recall at top $k$ ranks (R@k) to evaluate the performance of passage retrieval. 


\subsection{Comparison Methods}
To demonstrate the effectiveness of our model, we adopt several state-of-the-art document retrieval models for comparison as follows.

\paratitle{Sparse Retrieval Models}
We first compare our model with sparse passage retrieval models, including traditional retriever BM25~\cite{yang2017anserini}, and three retrievers enhanced by neural networks, DeepCT~\cite{dai2019deepct},  GAR~\cite{mao2020generation} and COIL~\cite{gao2021coil}. 

\paratitle{Dense Retrieval Models}
We compare with several dense passage retrieval models, including DPR~\cite{karpukhin2020dense}, ANCE~\cite{xiong2020approximate}, ColBERT~\cite{khattab2020colbert}, NPRINC~\cite{lu2020nprinc}, RocketQA~\cite{qu2021rocketqa}, PAIR~\cite{ren2021pair} and RocketQAv2~\cite{renrocketqav2}. For PAIR and RocketQAv2, we initialize them with BERT$_{\rm base}$ and reproduce their results on TQA. And we also compare with some pre-training methods, including Condenser~\cite{gao2021condenser} and coCondenser~\cite{gao2021unsupervised}.

\subsection{Implementation Details}\label{sec:id}
For fair comparison with baseline models, the dual-encoder is initialized with ERNIE-2.0 base \cite{zhang2019ernie} 
which is a BERT-like model with 12-layer transformers 
on MSMARCO and NQ, and is initialized with BERT$_{\rm base}$ on TQA. Our initial dual-encoder $M_{de}$ is trained with the batch sizes of $2048 \times 2$ on MSMARCO, and $256 \times 4$ on NQ and TQA. The number of epoch, the rate of linear scheduling warm-up and the learning rate are set to 10, 0.1 and 2e-5, respectively on all datasets.

While jointly training dual-encoder and GNN, we set the learning rate to 2e-6 and 5e-5 for dual-encoder and GNN, respectively, the batch size to $1024 \times 4$ and the number of epoch to 5. Since different network structures are suitable for different learning rate, it is necessary to set them different. We retrieve the top-25 candidate passages for each query by $M_{de}$ to create edge set $\mathcal{E}_{pq}$. For each epoch, we randomly select 5\% of the queries for training while others and their relevant edges for constructing graph. \begin{table}[t]
\centering
  \begin{tabular}{p{2.8cm}cccccccccc}
    \toprule
    Methods & R@5 & R@20 & R@100 \\
    \midrule
    BM25~(anserini) & - & 66.9 & 76.7 \\
    DPR (single) & - & 79.4 & 85.0  \\
    ANCE (single) & - &80.3 & 85.3 \\
    GAR & 73.1 & 80.4 & 85.7 \\
    PAIR & 76.3$^*$ & 82.4$^*$ & 86.9$^*$ \\
    Condenser &  - & 81.9 & 86.2 \\
    RocketQAv2 & 76.4$^*$ & 82.6$^*$ & 86.7$^*$ \\
    coCondenser & 76.8 & 83.2 & \textbf{87.3} \\
    \midrule
    \gmodel & \textbf{77.7} & \textbf{83.3} & 87.2 \\
  \bottomrule
\end{tabular}
  \caption{Experimental results on TriviaQA test set. The results of PAIR and RocketQAv2 are reproduced (marked with $^*$), while others are copied from the original paper. All dense retrieval models are initialized with BERT$_{\rm base}$. 
  }
  \label{tab:triviaR}
\end{table}
During the inference, we use the query encoder to predict query embeddings, and use FAISS~\cite{johnson2019billion} to index the dense representations of all passages via GNN-encoder.
We implement all experiments with the deep learning framework PyTorch on up to four NVIDIA Tesla A100 GPU (80GB memory).

\subsection{Experimental Results}
The detailed experimental results of passage retrieval tasks on MSMARCO and NQ are shown in Table~\ref{tab:mainR}, while the results of TQA are shown in Table~\ref{tab:triviaR}. We can observe that our model outperforms other fine-tuning methods by a large margin, especially on MRR@10 (+0.5\%), R@50 (+0.5\%) of MSMARCO, R@5 (+1.7\%), R@20 (+1.2\%) of NQ and R@5 (+1.3\%), R@20 (+0.7\%) of TQA. This phenomenon reflects that our model can build better query and passage representations to improve the ability of passage ranking at top ranks, which is due to our interaction between query embeddings and passage embeddings. When compared to pre-training methods like coCondenser, our model still performs better in general on all datasets. Note that pre-training methods are not comparable to our method in practice, but complementary work.

\subsection{Ablation Study}
\begin{table}[!t]
\centering
\begin{adjustbox}{max width=\linewidth}
    \begin{tabular}{lcccccccccc}
    \toprule
    Methods & R@5 & R@20 & R@100 \\
    \midrule
    \gmodel & \textbf{76.8} & \textbf{84.9} & \textbf{89.3} \\
    \midrule
    w/o GNN & 75.1 & 83.6 & 89.1 \\
    w/o MGT & 76.0 & 84.4 & 89.1 \\
    w/o filter gate &  76.3 & 84.6 & 89.2  \\
    w/o edge features & 76.5 & 84.7 & 89.2 \\
    w/ one layer & 76.3 & 84.5 & 89.2 \\
  \bottomrule
\end{tabular}
\end{adjustbox}
  \caption{Ablation study of different components of \gmodel on Natural Questions. 
  }
  \label{tab:ablation}
\end{table}

We perform an ablation study to investigate where the improvement mainly comes from. We only report the results on NQ which are shown in Table~\ref{tab:ablation}, while the results on MSMARCO and TQA are similar and omitted here due to limited space.

First, we use Eq.(\ref{eq:sim}$\sim$\ref{loss}) to train a dual-encoder without GNN on query-passage graph and compare it with our model to explore how much performance improvement is introduced by GNN.
The performance of our model notably drops in terms of R@5 and R@20 without GNN. 
It indicates that our model builds better query and passage embeddings by GNN interaction, which improves the performance at top ranks.

\begin{figure}
     \centering
     \includegraphics[width=1.0\columnwidth]{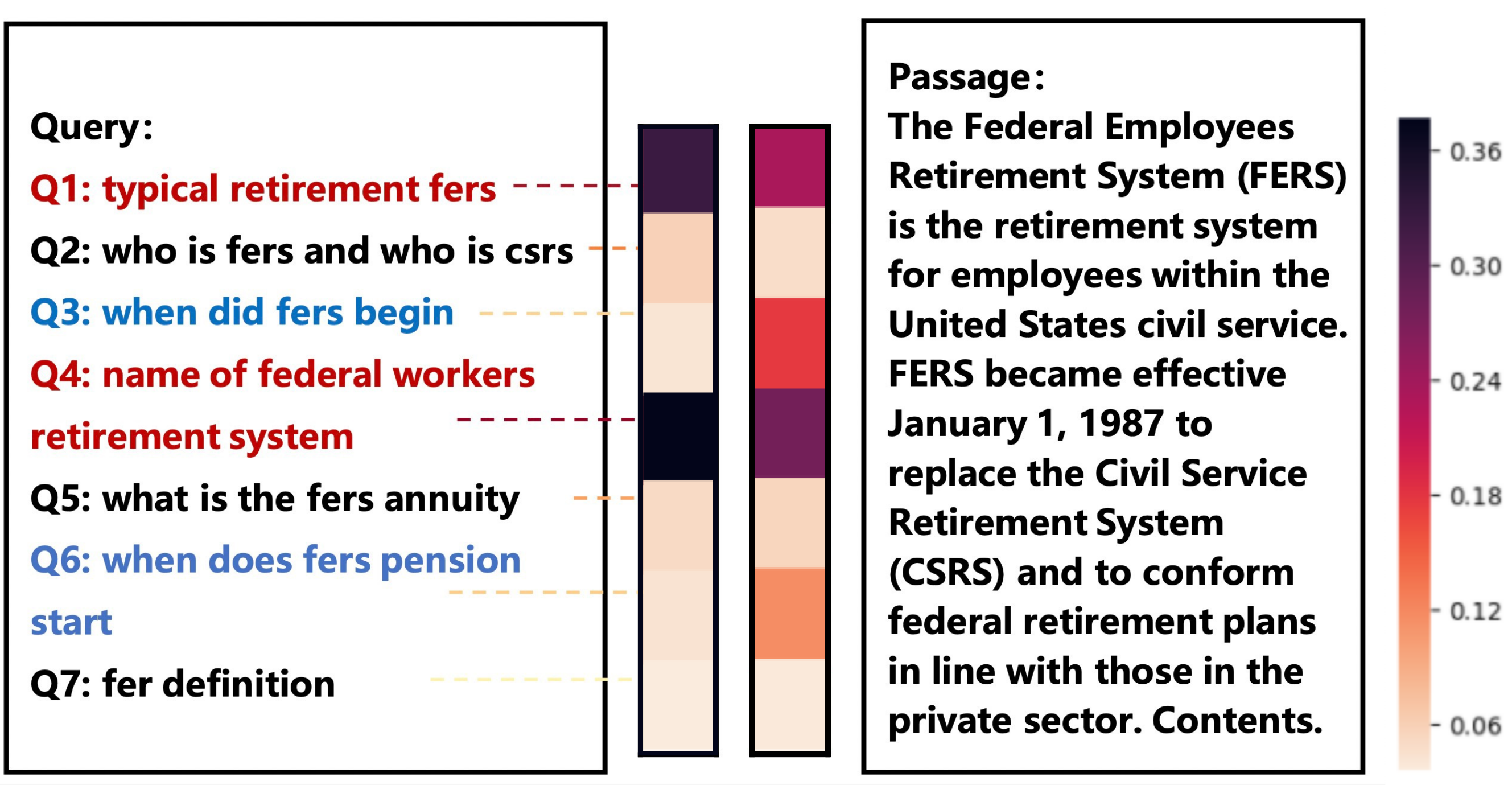}
     \caption{
         Attention scores in Eq.(\ref{eq:atention}) for w/o MGT (left) and w/ MGT (right). We illustrate a passage and queries connecting to it in $\mathcal{G}$, and mark labeled positive queries in training dataset and potentially relative queries in {\color{red}red} and {\color{blue}blue}, respectively.
     }
     \label{fig:attention}
\end{figure}

We use the whole training queries to both construct the graph and train \gmodel at the same epoch and drop edges between training queries and their positive passages to avoid information leakage like DGL~\cite{wang2019deep} instead of MGT Algorithm. 
However, it leads to a considerable drop in performance as shown in Table~\ref{tab:ablation}. We conclude the reason as in-batch negatives information leakage. 
As illustrated in Figure~\ref{fig:attention}, w/o MGT can lead to information leakage, because the passage embeddings will be more inclined to integrate the queries labeled positive in training dataset but ignore potentially relative queries \footnote{If query $q$ is supposed to have positive passage $p$, but for some reason it is not labeled, we consider $q$ to be a potentially relative query of $p$.} (as left attention scores show scores of labeled positive queries are much bigger than the others). 
However, our algorithm can solve this problem, due to queries for training are not used to construct graph in the same training epoch (as right attention scores show scores of potentially relative queries become larger). 

And then, we remove the filter gate in Eq.(\ref{eq:ipe}) and calculate query-interactive passage embeddings directly with a constant $\alpha$, \ie~$h_{p_i}' = \alpha \cdot \widetilde{h_{p_i}} + h_{p_i}$. For a more fair comparison, we searched for $\alpha$ from $0$ to $1$ by setting an equal interval to $0.1$, and release the best result in Table~\ref{tab:ablation} where $\alpha$ is set to $0.2$. However, the best result of replacing filter gate with constant $\alpha$ is still lower than that of filter gate. A potential reason is that the value of $\alpha$ should not be the same for different passages.

We also investigate whether the embeddings of cross encoder as edge features can effectively guide model to learn attention mechanism. As shown from results, the performance slightly drops when we remove the edge features. It indicates that GAT layer can learn how to select information from neighbors by only relying on passage embeddings and query embeddings, but the embeddings of cross encoder help it learn better.

We utilize one GAT layer instead of two layers to examine the effect of two-hop neighbors. As shown in Table \ref{tab:ablation} (w/ one layer), the performance of one layer drops compared to that of two layers. We think that the passages retrieved by a query which are also part of the query information, will complement the query embeddings. That is also the reason why two layers have better performance.

\subsection{Detailed Analysis}\label{DA}
Apart from the above illustration, we also implement detailed analysis on different settings of \modelnospace's training and efficiency.

\paratitle{The number of edges in Graph}
\begin{figure}
    \centering
    \includegraphics[width=1.0\columnwidth]{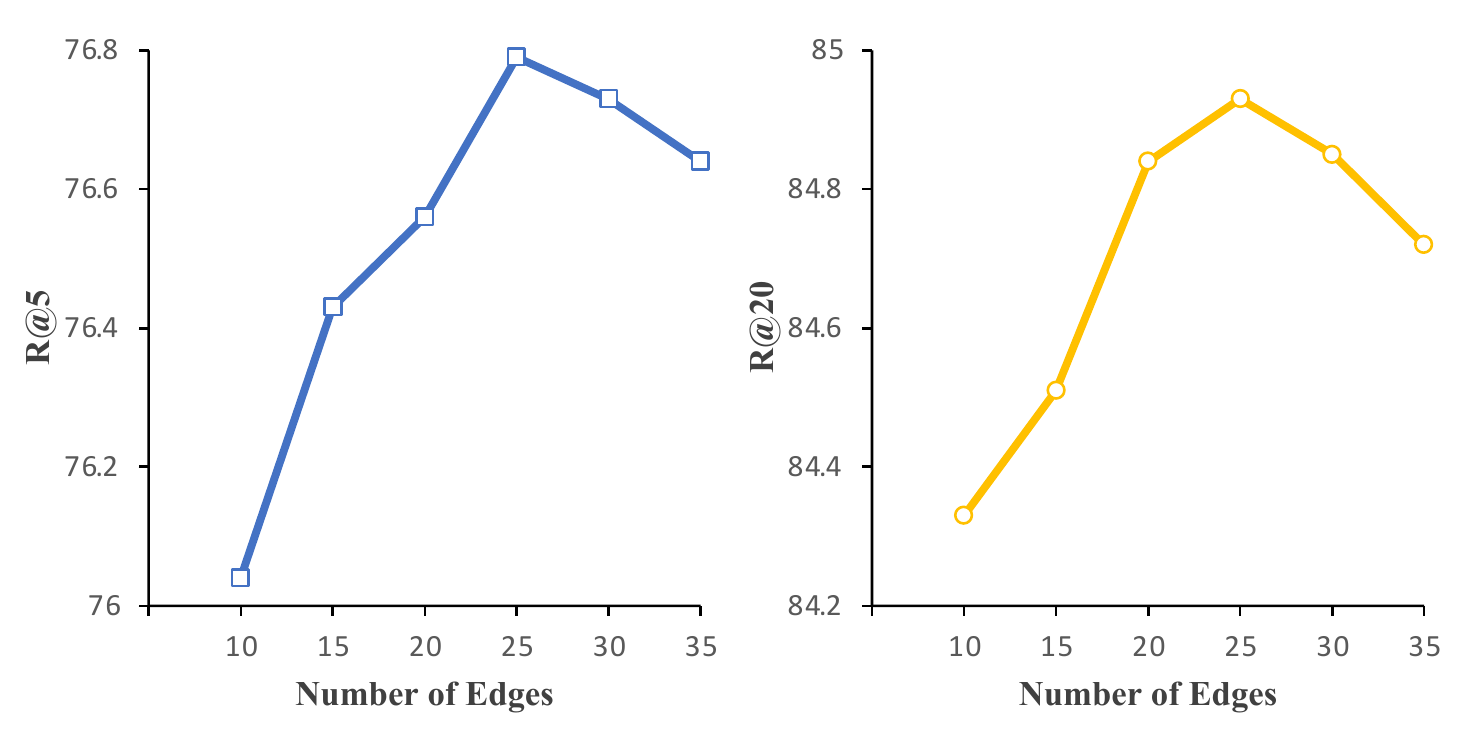}
    \caption{
        R@5 (left) and R@20 (right) results of passage retrieval on NQ with different numbers of edges per query node.
    }
    \label{fig:egde}
\end{figure}
For graph construction in section~\ref{3.2}, the more candidate passages are retrieved, the more edges there will be, which means that more queries connect to a passage and more query information could be incorporated into passage embeddings. 
However, Figure~\ref{fig:egde} indicates that more edges do not bring further performance improvements, because excessive edges might introduce noise that increases the difficulty of model learning and leads to performance drop.

\paratitle{Masked Ratio}
\begin{figure}
    \centering
    \includegraphics[width=1.0\columnwidth]{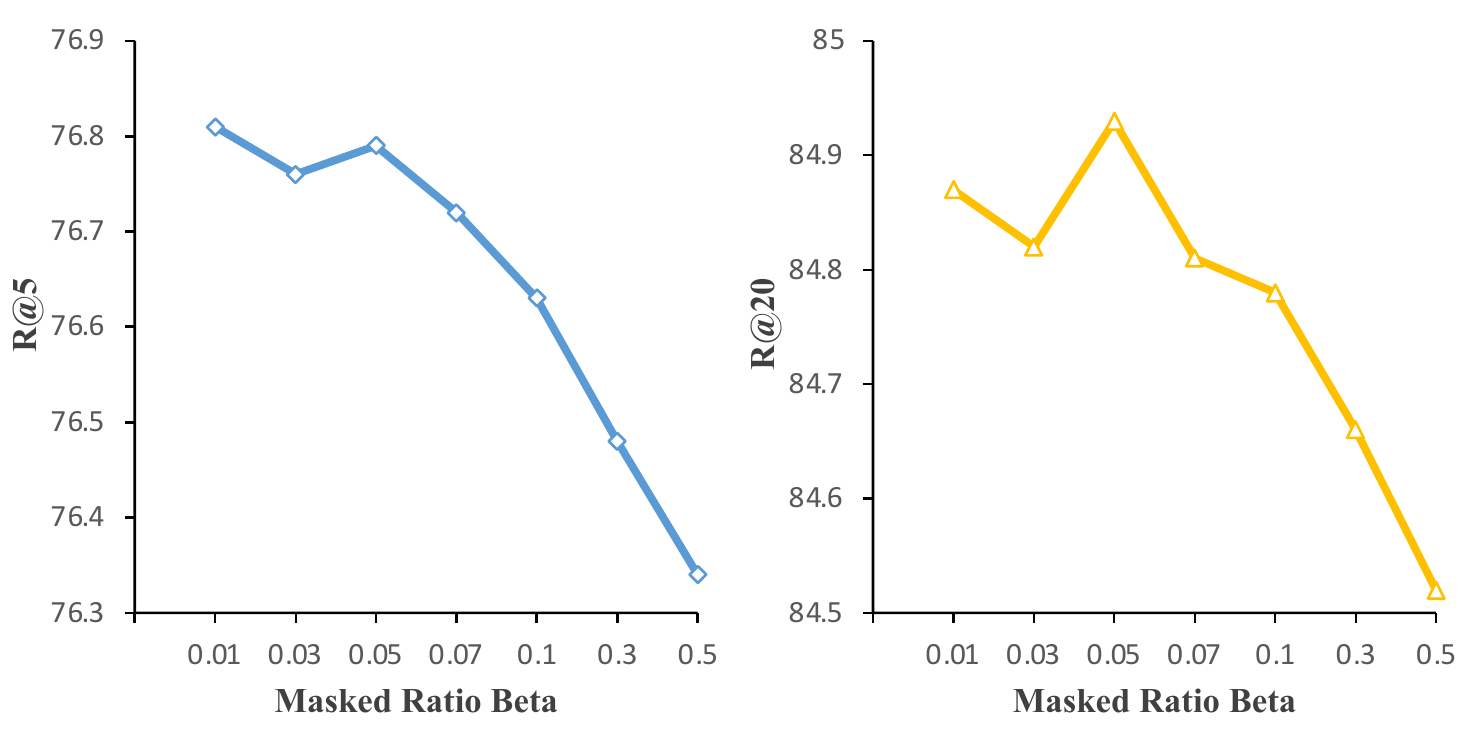}
    \caption{
        R@5 (left) and R@20 (right) results of passage retrieval on NQ with different masked ratios.
    }
    \label{fig:beta}
\end{figure}
In this part, we conduct an experiment to analyze the impact of the masked ratio on retrieval performance. 
As shown in Figure~\ref{fig:beta}, high masked ratio $\beta$ has poor performance, because it causes the discrepancy between the training graph and the inference graph~(the large $\beta$ means a few training queries used to construct graph in a training epoch). 
But it is also not necessary to set $\beta$ too small, since it will bring overhead for frequent constructing graph without any performance improvement~(the small $\beta$ means a few queries used to train which leads to more training epochs).

\begin{table}[t]
\centering
  \begin{tabular}{p{2.5cm}cccccccccc}
    \toprule
    Methods & Doc Encoding & Retrieval \\
    \midrule
    DPR & 0.41ms & 1.9ms \\
    ColBERT & 0.41ms & 90ms \\
    RocketQAv2 & 0.41ms & 1.9ms \\
    \gmodel & 0.46ms & 1.9ms \\
  \bottomrule
\end{tabular}
  \caption{Time cost of online retrieval and offline document encoding for Natural Questions test set.}
  \label{tab:effic}
\end{table}
\paratitle{Efficiency} We test the efficiency of our model on a single NVIDIA Tesla A100 80GB GPU for the NQ test set, and record the encoding time per document and retrieval time per query (including query encoding time), as shown in Table \ref{tab:effic}. Since we need to additionally compute query-interactive passage embeddings via GNN, the document encoding time is slightly longer than dual-encoder models. 
Since query-interactive passage embeddings can be pre-computed offline once obtained from \gmodel, we only need to encode query. Therefore we do not bring additional computation overhead to online retrieval (same as dual-encoder models).
\pdfoutput=1
\section{CONCLUSION}
In this paper, we introduced \gmodel for passage retrieval tasks and demonstrated its effectiveness on MSMARCO, Natural Questions and TriviaQA datasets. 
The existing dual-encoder architecture, although very efficient, ignores interaction during passage (query) encoding due to its independent architecture. 
Therefore we attempted to fuse query (passage) information into passage (query) representations via graph neural networks and maintain online efficiency of the dual-encoder. However, we may retrieve irrelevant passages for queries by dual-encoder when constructing the query-passage graph, which introduces noise in information propagating. Hence we utilize GAT layers and filter gate to reduce the noise, which are proved necessary by our various experiments. In the future, we will explore how to fuse more interaction information into GNN structure.
\pdfoutput=1
\section*{Limitations}
In this section, we will discuss the limitations of our work, which we consider as two major points: the requirement of more physical memory and utilizing cache passage embeddings as passage node features.

In both training and document encoding process of \gmodel, we need cache passage embeddings of $M_{de}$ ($m$ passage nodes) and embeddings of cross-encoder (($n\times k + m +n$)  edges), which should be calculated and stored in advance. It means that we need to store at least ($(n\times k + 2m + n) \times d$) floating-point numbers, where d is the dimension of BERT base. In practice, the number of passages $m$ is often very large, for example, the NQ dataset has about 20 million passages. Feature compression may be a good solution, but it may lead to performance drop.

As mentioned in Section \ref{3.4}, we utilize cache passage embeddings of $M_{de}$ as passage node features to compute $\widetilde{h_{q_i}}$ in Eq.(\ref{eq:gat1}) instead of recomputing them by passage encoder. It may be not the best approach for joint training dual-encoder and GNN, but it is a more practical way considering efficiency and hardware resources. We have tried to update cache passage embeddings, but it brings very little improvement and increases convergence difficulty.

\section*{Acknowledgements}
This work is supported by National Key R\&D Program of China (No. 2021YFC3340303) and National Natural Science Foundation
of China (NSFC Grant No. 62122089 and No. 61876196). We would also like to thank the anonymous reviewers for their helpful comments and suggestions.
\bibliography{anthology,custom}
\bibliographystyle{acl_natbib}

\newpage
\appendix
\pdfoutput=1
\begin{table*}[htp]
 \centering
  \begin{tabular}{p{0.5cm}p{3cm}p{3cm}p{9cm}}
    \toprule
    \#&Dev Query&Training Query&Relevant Passage\\
    \midrule
    1 & how long do items take that come from china & how long does a seller on ebay have to ship & The main risks I have encountered is that it takes about 10 days to 2 weeks for things to arrive from China to the US once they are mailed. eBay allows 30 days however. If a seller ships quickly, your items will arrive quickly and all will be well. However, be aware if your items do not arrive within that time frame. \\
  \bottomrule
\end{tabular}
  \caption{The example of MSMARCO retrieval result from \gmodel. We select a dev query which our model retrieves a positive passage at top-1 rank, and display a training query that have the same positive passage.}
\label{tab:examR}
\end{table*}

\section{Masked Graph Training Algorithm} \label{appendix:algorithm}
Algorithm \ref{alg:Framwork} is a more detailed and complete description of jointly training dual-encoder and GNN by our proposed Masked Graph Training Algorithm.

\begin{algorithm}[htp]
\caption{Masked Graph Training Algorithm} 
\label{alg:Framwork} 
\begin{algorithmic}[1]
\REQUIRE ~~ 
Training queries $Q$; Passages $P$; Dual-encoder ($E_Q(\cdot)$, $E_P(\cdot)$) initialized with $M_{de}$; Cross-encoder $M_{ce}$; Training data $C$.
\STATE Get cache passage embeddings: $h_{p_i}=E_P(p_i)$, and retrieve the top-k passages $P_i$ for each query $q_i$ by $M_{de}$ to get edge set $\mathcal{E}$.
\STATE Get edge features: $h_{x-y} = M_{ce}(x,y)$ for each edge $e(x,y)$ in $\mathcal{E}$.
\FOR{each epoch}
\STATE Split $Q$ into $Q_g$ and $Q_t$ by masked ratio $\beta$.
\STATE Use node sets $Q_g \cup P$ and their relevant edges to construct graph $\mathcal{G}$.
\FOR{each batch $Q_b \in Q_t$}
\STATE Get denoised positives and hard negatives of $Q_b$ from $C$: $P_b = P_b^+ \cup P_b^-$.
\STATE Utilize query embeddings recomputed by $E_Q(\cdot)$ and cache passage embeddings as node features and edge features to compute  $\widetilde{h_{q_i}}$ by Eq.(\ref{eq:gat1}) for each query which has an edge with the passage in $P_b$.
\STATE Utilize embeddings recomputed by dual-encoder as node features, $\widetilde{h_{q_i}}$ and edge features to compute $h_{p_i}'$ by  Eq.(\ref{eq:iqe}$\sim$\ref{eq:ipe}) for each passage in $P_b$.
\STATE Utilize $E_Q(q_i)$ and $h_{p_i}'$ to compute similarity and loss by Eq.(\ref{eq:gnnsim}) and Eq.(\ref{eq:gnnloss}).\STATE Update parameters of dual-encoder ($E_Q(\cdot)$, $E_P(\cdot)$) and GNN.
\ENDFOR
\ENDFOR

\end{algorithmic}
\end{algorithm}

\section{Case Study}
We also analyze the reasons why \gmodel outperforms RocketQAv2 by case study.
As Table~\ref{tab:examR} shows, we 
display the example of the MSMARCO top-1 retrieval results from our model which is not retrieved by RocketQAv2 to further illustrate how GNN integrates query information into passage embeddings effectively. 
We can observe that the passage are too long to be retrieved by the dev query, and we conclude the reasons as the fact that the long passage embeddings contain too much information to be focused on the key information.
Correspondingly, our model can incorporate the training query information into relevant passage embeddings, thus for a dev query that is similar to the training query, it is easier to retrieve this relevant passage.
For example, the dev query in Table~\ref{tab:examR} is similar to training query, so our model can easily retrieve the relevant passage and rank it at the first place.

\section{Data Statistics}
\begin{table}[!t]
\centering
\tabcolsep=0.12cm
\begin{adjustbox}{max width=\linewidth}
    \begin{tabular}{lcccc}
    \toprule
    \bf Datasest & \bf \#q in train & \bf \#q in dev & \bf \#q in test & \bf \#p \\
    \midrule
    MSMARCO & 502,939 & 6,980 & 6,837 & 8,841,823\\
    NQ & 79,168 & 8,757 & 3,610 & 21,015,324\\
    TQA &  78,785 & 8,837 & 11,313 & 21,015,324\\
  \bottomrule
\end{tabular}
\end{adjustbox}
  \caption{The statistics of datasets MSMARCO, NQ and TQA. Here, \#q and \#p denote the number of query in set and all passage.
  }
  \label{tab:data_statistics}
\end{table}
Table \ref{tab:data_statistics} shows the statistics of datasests MSMARCO, NQ and TQA. Following DPR \cite{karpukhin2020dense}, we discard the queries without golden passage for NQ and TQA.

\end{document}